\documentclass[aps,prl,preprint,superscriptaddress]{revtex4-2}

\usepackage{amsmath}
\usepackage{makecell}
\usepackage{graphicx}
\usepackage{dcolumn}
\usepackage{bm}
\usepackage{color}
\usepackage{siunitx}
\usepackage[colorlinks=true,linkcolor=black,citecolor=black,urlcolor=black]{hyperref}
\usepackage{csquotes}

\usepackage{textcomp}

\usepackage{color}
\begin{document}
	
	\title{Influence of twist angle on ultrafast charge separation in WS\textsubscript{2}-graphene heterostructures}
	
	
	
	\author{Niklas Hofmann}
	\affiliation{Department for Experimental and Applied Physics, University of Regensburg, 93053 Regensburg, Germany}
	
	\author{Leonard Weigl}
	\affiliation{Department for Experimental and Applied Physics, University of Regensburg, 93053 Regensburg, Germany}
	
	\author{Johannes Gradl}
	\affiliation{Department for Experimental and Applied Physics, University of Regensburg, 93053 Regensburg, Germany}
		
	\author{Stiven Forti}
	\affiliation{Center for Nanotechnology Innovation IIT@NEST, Istituto Italiano di Tecnologia, Pisa, Italy}

	\author{Domenica Convertino}
	\affiliation{Center for Nanotechnology Innovation IIT@NEST, Istituto Italiano di Tecnologia, Pisa, Italy}
	
	\author{Camilla Coletti}
	\affiliation{Center for Nanotechnology Innovation IIT@NEST, Istituto Italiano di Tecnologia, Pisa, Italy}
	
	\author{Isabella Gierz}
	\email[]{isabella.gierz@ur.de}
	\affiliation{Department for Experimental and Applied Physics, University of Regensburg, 93053 Regensburg, Germany}
	
	
	\date{\today}
	
	\begin{abstract}
	
	Van der Waals (vdW) heterostructures, formed by stacking two-dimensional materials, offer highly tunable electronic and optical properties, with the twist angle between layers acting as a critical tuning parameter. While its impact on moiré patterns, band structure, and correlated states is well-established, the influence of twist angle on ultrafast charge transfer remains controversial. Here, we employ time- and angle-resolved photoemission spectroscopy (trARPES) to directly probe ultrafast charge transfer in epitaxially grown WS\textsubscript{2}-graphene heterostructures with twist angles of 0$^{\circ}$ and 30$^{\circ}$. Upon photoexcitation at $\hbar\omega = 3.1\,\mathrm{eV}$, we observe efficient charge separation at 0$^{\circ}$, while at 30$^{\circ}$, electron and hole transfer occur at similar rates. Our results highlight the crucial role of the twist angle in controlling charge separation efficiency, offering valuable insights for designing vdW heterostructures for applications in photovoltaics and optoelectronics. 
	\end{abstract}
	
	
	\maketitle

\section{Introduction}

Van der Waals (vdW)  heterostructures are artificial materials created by stacking different 2D solids on top of each other, held together by weak vdW forces. This allows for the design of materials with highly tunable electronic and optical properties, independent of lattice matching constraints. In this context, the twist angle serves as a powerful tuning knob as it influences the periodicity of the moiré potential \cite{Shabani2021} and thereby moiré excitons \cite{Tran2019} and phonons \cite{Lin2018}, band structure \cite{Ohta2012, Ulstrup2020}, and electron-electron interactions \cite{Kerelsky2019}. At specific magic twist angles, such as $\sim1.1^{\circ}$ in twisted bilayer graphene, vdW heterostructures develop flat electronic bands resulting in the emergence of strongly correlated states \cite{Cao2018, Lisi2021}.

The twist angle is also expected to influence ultrafast charge transfer processes across different vdW interfaces \cite{Jin2018, DalConte2020, Jiang2021} which are crucial for applications in light harvesting and detection. In this context, literature provides a variety of experimental observations. A coherent microscopic understanding, however, seems to be missing. For example, the rate of ultrafast hole transfer from MoS\textsubscript{2} to WS\textsubscript{2} was found to be independent of the interlayer twist angle \cite{Ji2017, Zhu2022}, while the rate of ultrafast electron transfer in WSe\textsubscript{2}-WS\textsubscript{2} \cite{Merkl2019} and MoS\textsubscript{2}-WSe\textsubscript{2} heterostructures \cite{Zimmermann2021} showed a strong variation with twist angle. Recent ultrafast electron diffraction experiments suggest that charge transfer from MoS\textsubscript{2} to graphene gets slower with increasing twist angle \cite{Luo2021}. Finally, efficient charge separation was found to occur in WS\textsubscript{2}-graphene heterostructures with a twist angle of 0$^{\circ}$ \cite{Aeschlimann2020b, Krause2020, Hofmann2023} while no indication for charge separation was found in MoS\textsubscript{2}-graphene heterostructures with a twist angle of 30$^{\circ}$ \cite{Ulstrup2016}.

To shed light onto this issue, we use time- and angle-resolved photoemission spectroscopy (trARPES) to provide a direct view on ultrafast charge transfer processes in epitaxially grown WS\textsubscript{2}-graphene heterostructures with twist angles of 0$^{\circ}$ and 30$^{\circ}$. Following photoexcitation at $\hbar\omega=3.1\,\mathrm{eV}$ we confirm the occurrence of efficient charge separation for he\-te\-ro\-structures with a twist angle of 0$^{\circ}$. For a twist angle of 30$^{\circ}$, no charge separation is observed, indicating that electron and hole transfer occur at similar rates. Our findings provide fundamental insights into controlling charge separation efficiency through the twist angle, enabling the rational design of vdW heterostructures for advanced applications in photovoltaics and optoelectronics.

\section{Methods}

\begin{figure}
		\includegraphics[width = .4\columnwidth]{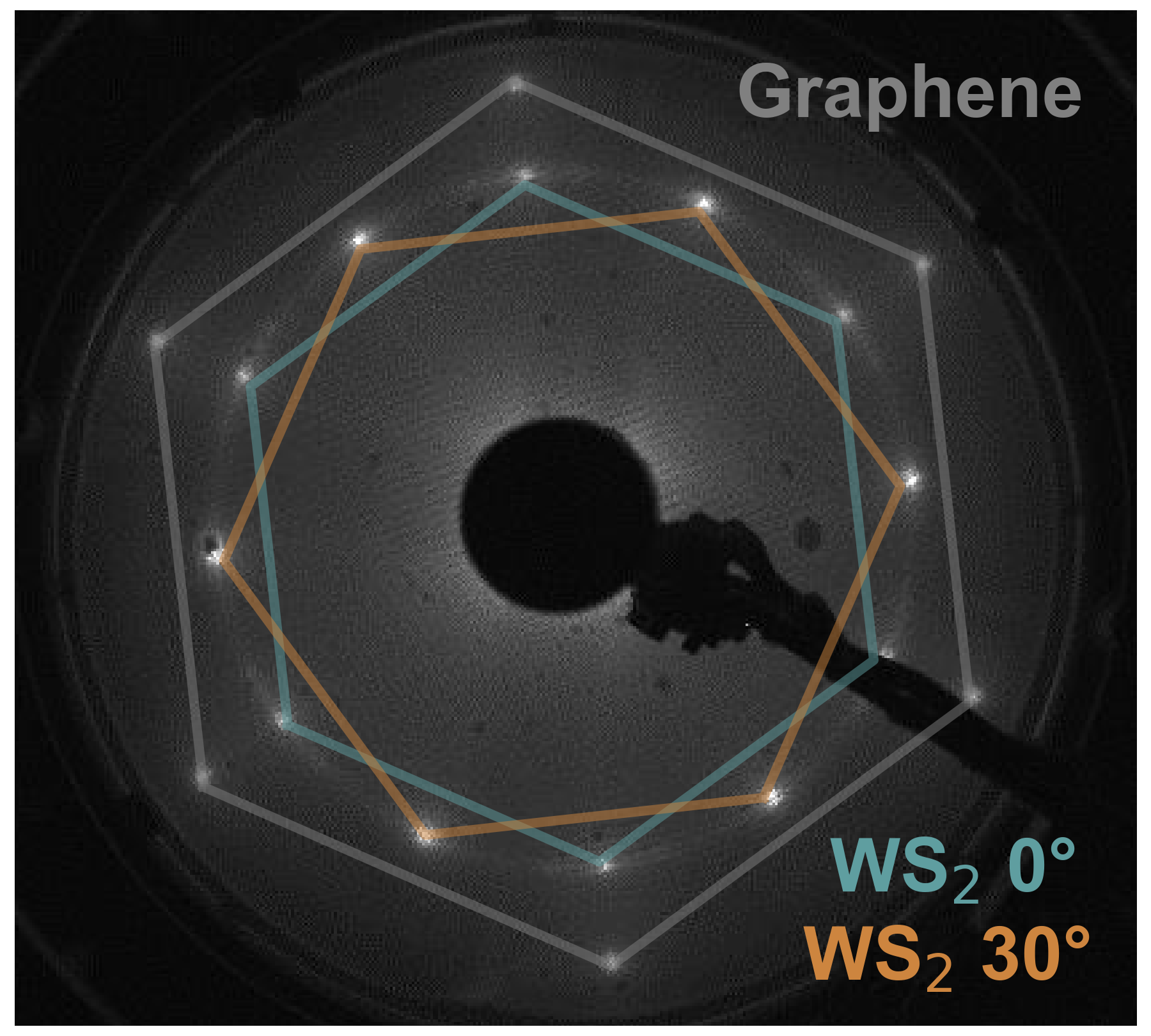}
		\caption{\textbf{LEED image of WS\textsubscript{2}-graphene heterostructure.} LEED image taken with an electron energy of $E_\mathrm{kin}=68\,\mathrm{eV}$. The grey hexagon marks the diffraction spots of graphene. Blue and orange hexagons indicate the diffraction spots from WS\textsubscript{2} islands with twist angles of 0$^{\circ}$ and 30$^{\circ}$ relative to the graphene layer, respectively.}
		\label{figure1}
	\end{figure}

WS\textsubscript{2} islands were grown by chemical vapor deposition from solid precursors \cite{Rossi2016, Forti2017} on top of hydrogen-intercalated epitaxial graphene on SiC(0001) \cite{Emtsev2009, Riedl2009}. The WS\textsubscript{2} islands exhibit a triangular shape with edge lengths ranging from $300\,\mathrm{nm}$ to $700\,\mathrm{nm}$ \cite{Krause2020, Hofmann2023} and grow mainly with twist angles of either 0$^{\circ}$ or 30$^{\circ}$ with respect to the underlying graphene layer, as demonstrated using low-energy electron diffraction (LEED) in Fig.~\ref{figure1}.

TrARPES experiments were performed at a repetition rate of $1\,\mathrm{kHz}$ using $3.1\,\mathrm{eV}$ pump pulses with a fluence of $0.4\,\mathrm{mJ}/\mathrm{cm}^2$ and $21.7\,\mathrm{eV}$ probe pulses. ARPES spectra were recorded with a hemispherical analyzer that collects photoelectrons from the $300\,\mathrm{\mu m}$-sized area on the sample surface illuminated by the probe pulse, thus averaging over many different WS\textsubscript{2} islands. The temporal and energy resolutions for the measurements presented in this work were $200\,\mathrm{fs}$ and $350\,\mathrm{meV}$, respectively.

\section{Results}

\begin{figure}
		\includegraphics[width = \columnwidth]{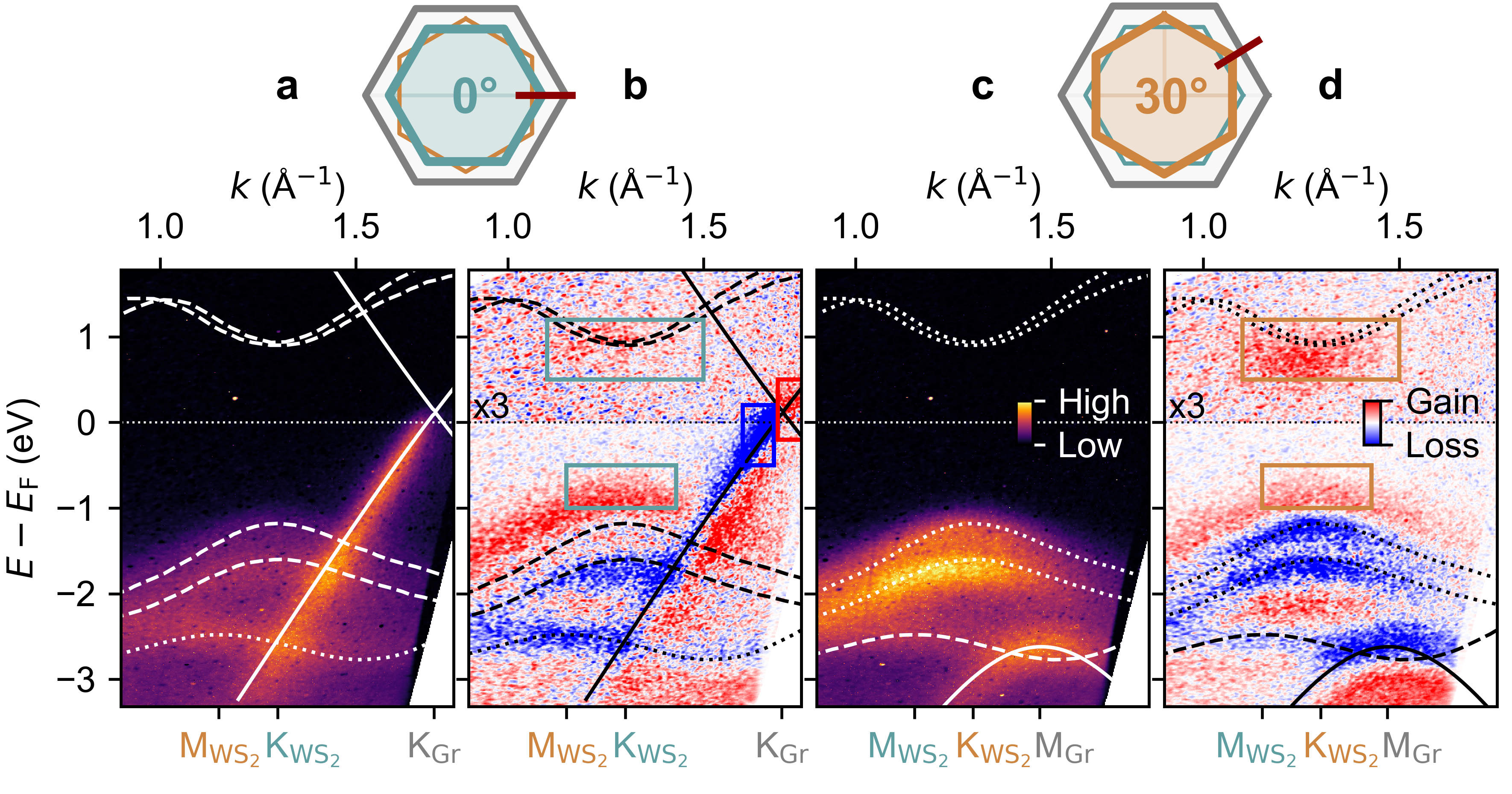}
		\caption{\textbf{TrARPES snapshots of WS\textsubscript{2}-graphene heterostructure.} a) Equilibrium ARPES image measured along the $\Gamma$K-direction of graphene (see inset, red line indicates field of view). b) Pump-induced changes of the spectrum from a) at a pump-probe delay of $310\,\mathrm{fs}$ after excitation with $\hbar\omega_\mathrm{pump}=3.1\,\mathrm{eV}$ with a fluence of $F=0.4\,\mathrm{mJ}/\mathrm{cm}^2$. c), d) Same as a), b) but along the $\Gamma$M-direction of graphene. Dashed (dotted) lines are guides to the eye indicating the equilibrium band structures of WS\textsubscript{2} \cite{Zeng2013} with a twist angle of 0$^{\circ}$ (30$^{\circ}$). Continuous lines mark graphene bands \cite{Wallace1947}. Colored boxes indicate the areas over which the photocurrent was integrated to yield the pump-probe traces displayed in Fig.~\ref{figure3}.}
	\label{figure2}
	\end{figure}

Figure \ref{figure2} shows the equilibrium ARPES spectra of a WS\textsubscript{2}-graphene heterostructure along the $\Gamma$K- and $\Gamma$M-directions of the hexagonal Brillouin zone of graphene in panels a and c, respectively, together with the pump-induced changes of the spectra at a pump-probe delay of $310\,\mathrm{fs}$ in panels b and d. The Brillouin zones of graphene and WS\textsubscript{2} with twist angles of 0$^{\circ}$ and 30$^{\circ}$ are shown in the insets together with a red line indicating the direction along which the band structure was measured. Dashed (dotted) lines are guides to the eye indicating the equilibrium band structures of WS\textsubscript{2} \cite{Zeng2013} with a twist angle of 0$^{\circ}$ (30$^{\circ}$). Continuous lines mark graphene bands \cite{Wallace1947}. The pump-probe signal contains signatures of the non-equilibrium carrier distribution as well as band shifts due to transient band gap renormalization and charging of the layers as previously discussed in \cite{Aeschlimann2020b, Krause2020, Hofmann2023}.

To analyze this rich pump-probe signal in detail, we start by integrating the counts over the areas marked by the colored boxes in Fig.~\ref{figure2}. Figure \ref{figure3}a shows the gain above the Fermi level and the loss below the Fermi level in the Dirac cone of graphene as a function of pump-probe delay as extracted from the red and blue boxes in Fig.~\ref{figure2}a. We find a short-lived gain (exponential lifetime of $0.12\pm0.03\,\mathrm{ps}$) and a long-lived loss (exponential lifetime of $2.2\pm0.2\,\mathrm{ps}$) in good agreement with our previous results \cite{Aeschlimann2020b, Krause2020, Hofmann2023}. The non-equilibrium population of the WS\textsubscript{2} conduction band (CB) is shown in Fig.~\ref{figure3}b for WS\textsubscript{2} islands with a twist angle of 0$^{\circ}$ (turquoise) and 30$^{\circ}$ (orange). We find that both pump-probe traces decay with an exponential lifetime of $\sim1.2\,\mathrm{ps}$ independent of the twist angle. The pump-probe signal in Fig.~\ref{figure3}c shows the gain above the equilibrium position of the WS\textsubscript{2} valence band (VB) for WS\textsubscript{2} islands with a twist angle of 0$^{\circ}$ (turquoise) and 30$^{\circ}$ (orange). Here, the 0$^{\circ}$ and 30$^{\circ}$ islands show a strikingly different behavior. The pump-probe trace of the 0$^{\circ}$ islands is found to decay with a single-exponential decay time of $1.5\pm0.1\,\mathrm{ps}$, whereas the 30$^{\circ}$ islands exhibit a double-exponential decay with decay times of $\tau_1=0.17\pm0.05\,\mathrm{ps}$ and $\tau_2=2.0\pm0.5\,\mathrm{ps}$, respectively. Details about the exponential fits are provided in the Supplemental Material.

In Fig.~\ref{figure4}a we present transient peak positions of the WS\textsubscript{2} VB and CB for both 0$^{\circ}$ and 30$^{\circ}$ islands. Details about the data fitting are provided in the Supplemental Material. We find that the bands of the 0$^{\circ}$ islands shift up, while the bands of the 30$^{\circ}$ islands shift down. The transient band gap (see Fig.~\ref{figure4}b) is obtained by subtracting the VB position from the CB position. Both kinds of WS\textsubscript{2} islands show a transient reduction of the band gap that exponentially recovers on a timescale of a few picoseconds. After removing the contribution of the transient band gap renormalization (subtracting $\Delta E_\mathrm{gap}/2$ from the transient position of the CB and adding $\Delta E_\mathrm{gap}/2$ to the transient position of the VB, see \cite{Krause2020, Hofmann2023}), we find that the bands of the 0$^{\circ}$ WS\textsubscript{2} islands exhibit a transient upshift of $169\pm8\,\mathrm{meV}$, while the bands of the 30$^{\circ}$ WS\textsubscript{2} islands exhibit a transient downshift of $40\pm2\,\mathrm{meV}$ with exponential lifetimes of a few picoseconds (see Fig.~\ref{figure4}c).

The transient position of the Dirac cone  at the K-point shown in turquoise in Fig.~\ref{figure4}d was determined as described in detail in the Supplemental Material. In good agreement with our previously published results on 0$^{\circ}$ islands \cite{Aeschlimann2020b, Krause2020, Hofmann2023}, the graphene Dirac cone is found to shift down by $83\pm3\mathrm{meV}$ with an exponential lifetime of $1.3\pm0.1\,\mathrm{ps}$. The transient position of the graphene $\pi$-band at M (orange in Fig.~\ref{figure4}d, details are provided in the Supplemental Material) is found to shift down by $68\pm2\,\mathrm{meV}$ with an exponential lifetime of $1.9\pm0.2\,\mathrm{ps}$ similar to what we observed at the K-point.

Our main findings can be summarized as follows: (1) The population dynamics of the WS\textsubscript{2} CB and the transient band gap renormalization are independent of the twist angle. (2) The gain above the equilibrium position of the WS\textsubscript{2} VB follow a single-exponential decay for a twist angle of 0$^{\circ}$, and a double-exponential decay for a twist angle of 30$^{\circ}$.  (3) The WS\textsubscript{2} VB exhibits an up-shift for a twist angle of 0$^{\circ}$ and a down-shift for a twist angle of 30$^{\circ}$.

\section{Discussion}

Resonant excitation of the A-exciton of WS\textsubscript{2} at $\hbar\omega_\mathrm{pump}=2.0\,\mathrm{eV}$ has been shown to be followed by ultrafast hole transfer into the graphene layer, resulting in a charge-separated state with a lifetime on the order of one picosecond \cite{Aeschlimann2020b, Krause2020, Zhou2021, Hofmann2023}. The following microscopic picture for ultrafast charge transfer emerged: Despite the weak vdW interaction between WS\textsubscript{2} and graphene, density functional theory calculations showed that the two layers hybridize, resulting in wave functions that are delocalized over both layers (\enquote{charge transfer states}) \cite{Hofmann2023}, enabling efficient charge transfer between the layers \cite{Long2016, Zheng2017, Li2017}. Electrons and holes then relax to the Fermi level located inside the Dirac cone of graphene by the emission of WS\textsubscript{2} A\textsubscript{1g} and graphene E\textsubscript{2g} phonons \cite{Liu2021}. Hole transfer is faster than electron transfer because the tunneling matrix element for holes is bigger than the one for electrons and the potential barrier for holes is smaller than the one for electrons \cite{Krause2020}. In addition to this intrinsic tunneling channel, tunneling can also occur through sulphur vacancies \cite{Hernangomez-Perez2023, Gradl2026}, a common defect of monolayer WS\textsubscript{2} \cite{Schuler2019}. Note that, in the present study, the WS\textsubscript{2}-graphene heterostructures were excited at $\hbar\omega_\mathrm{pump}=3.1\,\mathrm{eV}$ rather than at the A-exciton resonance. This elevated excitation energy opens an additional, highly efficient intrinsic charge-transfer channel for holes in the WS\textsubscript{2} valence band \cite{Hofmann2025b}.

\begin{figure}
		\includegraphics[width = \columnwidth]{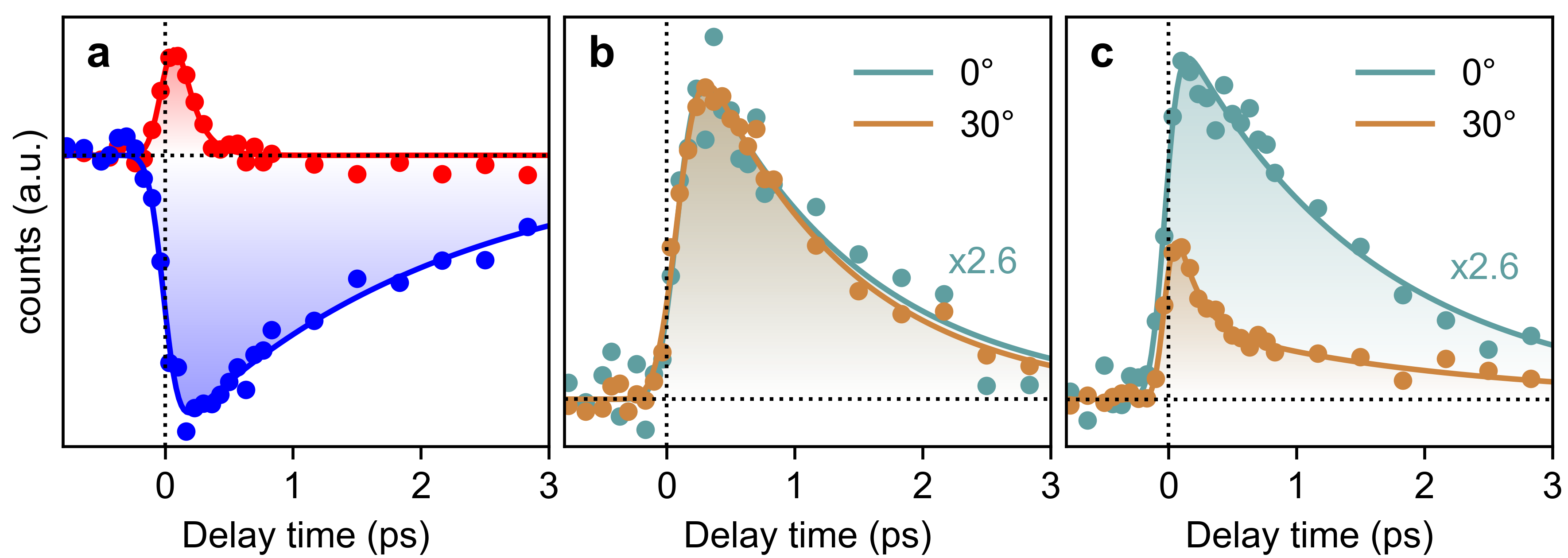}
		\caption{\textbf{Population dynamics.} a) Gain above the Fermi level (red) and loss below the Fermi level (blue) inside the Dirac cone as obtained by integrating the counts over the areas marked by the red and blue box in Fig.~\ref{figure2}b. b) Transient population of the WS\textsubscript{2} CB at K for both 0$^{\circ}$ and 30$^{\circ}$ islands. c) Gain above the equilibrium position of the WS\textsubscript{2} VB for both 0$^{\circ}$ and 30$^{\circ}$ islands. Data points in c) and d) were obtained by integrating the counts over the areas marked by the respective boxes in Figs. \ref{figure2}b and d. Continuous lines are exponential fits as described in the Supplemental Material. Vertical dashed lines indicate temporal overlap of pump and probe pulses, horizontal dashed lines indicate the equilibrium levels.}
	\label{figure3}
	\end{figure}

\begin{figure}
		\includegraphics[width = \columnwidth]{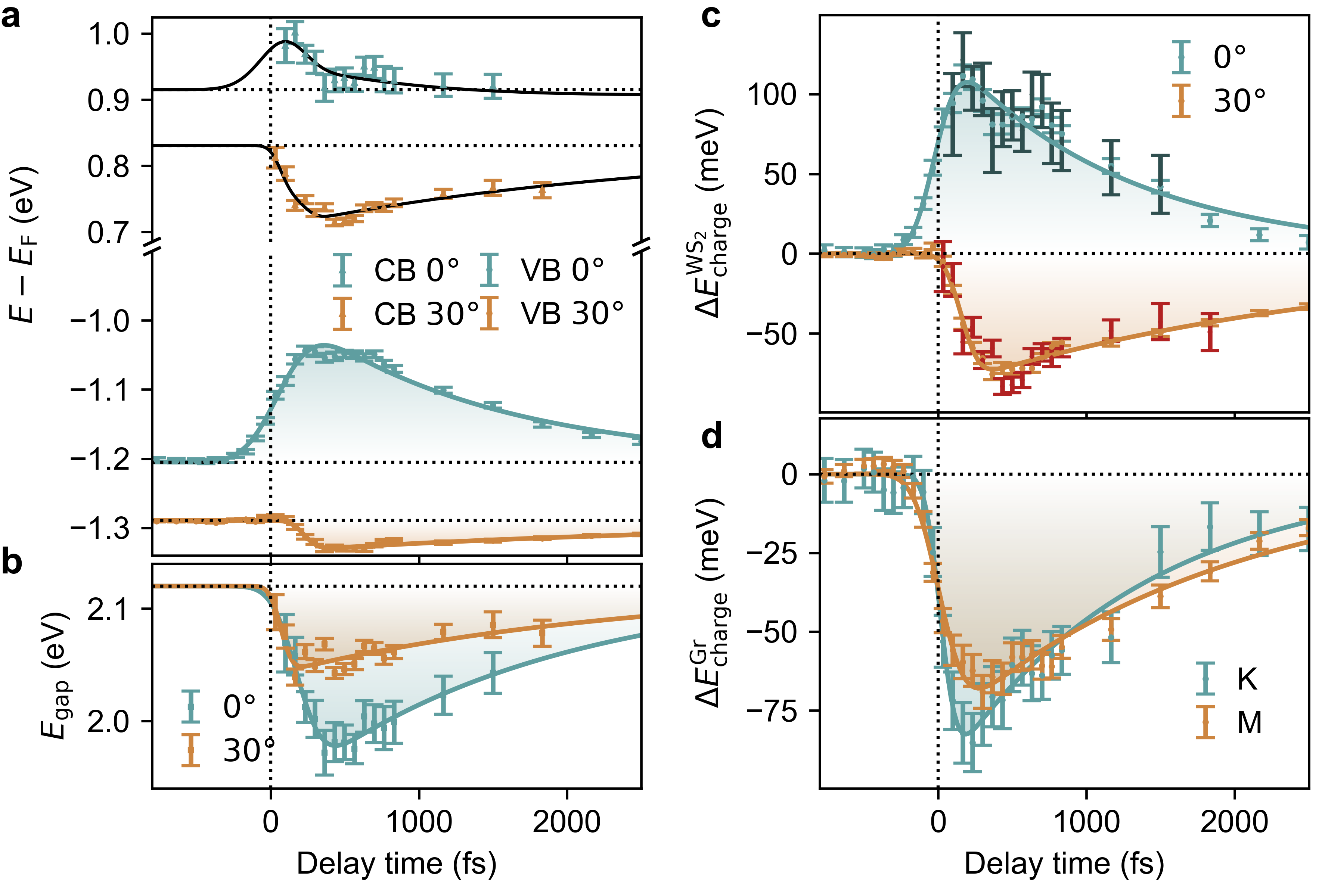}
		\caption{\textbf{Transient band structure.} a) Transient band positions of WS\textsubscript{2} CB (top) and VB (bottom). b) Transient band gap. c) Charging shift of WS\textsubscript{2}. Light (dark) data points belong to the WS\textsubscript{2} VB (CB). d) Charging shift of graphene. Colored lines are exponential fits as described in the Supplemental Material. Black lines in a) are calculated from the fits in b) and c).}
	\label{figure4}
	\end{figure}

The data for the 0$^{\circ}$ WS\textsubscript{2} islands as well as for the graphene $\pi$-band are in perfect agreement with this scenario \cite{Aeschlimann2020b, Krause2020, Hofmann2023}. Here, ultrafast charge separation is responsible for the asymmetric population dynamics inside the Dirac cone in Fig.~\ref{figure3}a and the charging shifts in Fig.~\ref{figure4}c and d, where excess electrons (holes) inside the WS\textsubscript{2} (graphene) layer reduce (increase) the binding energy of the WS\textsubscript{2} (graphene) bands. The transient reduction of the WS\textsubscript{2} band gap in Fig.~\ref{figure4}b is caused by light-induced free carriers that efficiently screen the Coulomb interaction \cite{Chernikov2015}. The gain above the equilibrium position of the WS\textsubscript{2} valence band in Fig.~\ref{figure3}c contains contributions from a transient broadening and upshift of the WS\textsubscript{2} valence band.

As shown in Figs. \ref{figure3}c and \ref{figure4}c, the 30$^{\circ}$ WS\textsubscript{2} islands exhibit behavior distinct from their 0$^{\circ}$ counterparts. Unlike the 0$^{\circ}$ WS\textsubscript{2} islands, which become negatively charged, the 30$^{\circ}$ islands acquire a positive charge, matching that of the graphene layer (Figs. \ref{figure4}c,d). This suggests that charge separation is unlikely to account for our observations. Furthermore, Fig.~\ref{figure3}b reveals that electrons leave the conduction band at the same rate for both twist angles. The lack of charge separation in the 30$^{\circ}$ islands therefore implies that the hole transfer rate must be comparable to the electron transfer rate. We speculate that the transient downshift of the WS\textsubscript{2} bands observed in Fig.~\ref{figure4}c may result from a back-gate effect induced by the positively charged graphene layer. Finally, the shape of the pump-probe signal in Fig.~\ref{figure3}c can be attributed to two competing effects: instantaneous band broadening and a band downshift with slower rise time (see Fig.~\ref{figure4}a). 

In summary, our results indicate that for a twist angle of 0$^{\circ}$, hole transfer is significantly faster than electron transfer, whereas for a twist angle of 30$^{\circ}$, electron and hole transfer occur at comparable rates.

These observations are in good agreement with previous trARPES data on MoS\textsubscript{2}-graphene heterostructures with a twist angle of 30$^{\circ}$ that also show no indications of charge separation \cite{Ulstrup2016}. Further, Ref. \cite{Luo2021} reports that charge transfer from MoS\textsubscript{2} to graphene gets slower with increasing twist angle. While a direct comparison with our trARPES data is difficult as Ref. \cite{Luo2021} does not distinguish between electron and hole transfer, this finding is consistent with our observation that hole transfer is faster for a twist angle of 0$^{\circ}$ than for a twist angle of 30$^{\circ}$.

In an attempt to understand the link between the twist angle and the presence or absence of charge separation, we now compare our results to previously published theoretical work that discusses MoS\textsubscript{2}-graphene \cite{Ebnonnasir2014} and WS\textsubscript{2}-graphene heterostructures \cite{Kleiner2024} with twist angles of 0$^{\circ}$ and 30$^{\circ}$. The most obvious difference between the band structures corresponding to the two twist angles is that for a twist angle of 0$^{\circ}$ the K-points of both the transition metal dichalcogenide (TMDC) and the graphene layer are folded onto the K-point of the mini Brillouin zone of the heterostructure, while for a twist angle of 30$^{\circ}$ the graphene K-point is folded onto the $\Gamma$-point of the mini Brillouin zone. Further, both references report different band alignments for twist angles of 0$^{\circ}$ and 30$^{\circ}$. For a twist angle of 0$^{\circ}$ the graphene Dirac point is located close to the center of the TMDC gap at K. For a twist angle of 30$^{\circ}$ the graphene Dirac point moves closer to the TMDC conduction band. We also observe this trend in our data (see differences in binding energy for negative delays in Fig.~\ref{figure4}a). The measured differences in band alignment are, however, much smaller than predicted theoretically. Ref. \cite{Ebnonnasir2014} further predicts a transition from direct to indirect MoS\textsubscript{2} band gap when changing the twist angle from 0$^{\circ}$ to 30$^{\circ}$ that is difficult to verify based on the measured band structures shown in Fig.~\ref{figure2}.

Ref. \cite{Kleiner2024} also plots the orbital composition of the states in the band structure, revealing the presence of charge transfer states with delocalized wave functions for both twist angles. Based on their calculations we can estimate the heights of the energy barriers that electrons and holes need to overcome in order to reach the closest charge transfer state. We estimate $\Delta E_\mathrm{CB}^{0^{\circ}}\sim470\,\mathrm{meV}$ and $\Delta E_\mathrm{VB}^{0^{\circ}}\sim170\,\mathrm{meV}$ for a twist angle of 0$^{\circ}$ and $\Delta E_\mathrm{CB}^{30^{\circ}}\sim180\,\mathrm{meV}$ and $\Delta E_\mathrm{VB}^{30^{\circ}}\sim500\,\mathrm{meV}$ for a twist angle of 30$^{\circ}$. Although the absolute values should be interpreted with caution due to the artificial strain applied to the graphene layer in the calculations, the estimated energy barriers suggest that increasing the twist angle from 0$^{\circ}$ to 30$^{\circ}$ should accelerate electron transfer and slow down hole transfer --- consistent with the experimental trend for holes but not for electrons. We would like to point out that tunneling of electrons through sulphur vacancies  \cite{Hernangomez-Perez2023, Gradl2026} is expected to be independent of twist angle as the sulphur vacancy states are delocalized over the full Brillouin zone. Thus, the absence of any twist angle dependence in Fig.~\ref{figure3}b would be consistent with a scenario where electron transfer is dominated by sulphur vacancies in this particular sample.

Estimating the magnitude of the tunneling matrix element from the calculations in \cite{Kleiner2024} is challenging, since the hybridization-induced avoided crossings are not distinctly resolved in the band structure. However, \cite{Kleiner2024} predicted that the interplay between interlayer and intralayer excitons strongly varies with twist angle. Because of this, both the absorption and the charge transfer between the layers were found to change. More precisely, efficient charge separation with electrons localized in the WS\textsubscript{2} layer and holes localized in the graphene layer was found to occur for a twist angle of 30$^{\circ}$ but not for 0$^{\circ}$ \cite{Kleiner2024}. Further, \cite{Liu2022} combined time-dependent density functional theory with nonadiabatic molecular dynamics to predict ultrafast charge separation in WS\textsubscript{2}-graphene heterostructures with a twist angle of 30$^{\circ}$. These predictions conflict with both our own trARPES measurements \cite{Aeschlimann2020b, Krause2020, Hofmann2023} and previous work by others \cite{Ulstrup2016}, which consistently show ultrafast charge separation at a twist angle of 0$^{\circ}$ but not at 30$^{\circ}$. A microscopic understanding of the experimental data clearly requires further theoretical investigations.

While band structures and charge transfer states in the model discussed above are material specific, the underlying concept is more general. Therefore, we provide a short discussion on the influence of the twist angle on ultrafast charge transfer in vdW heterostructures consisting of two different monolayer TMDCs. There, Ref. \cite{Zheng2017} predicts a strong hybridization between the bands of the individual layers in the VB at $\Gamma$. Since the $\Gamma$ points of the two layers coincide for any twist angle, the hole transfer rate should remain unaffected by twist angle. For the CB, however, Ref. \cite{Zheng2017} predicts hybridization at momenta $k>0$ only. Twisting the two layers with respect to each other then increases the momentum offset between the bands of the individual layers for any non-zero momentum away from $\Gamma$. Hence, hybridization and thus electron transfer is expected to be highly sensitive to the twist angle between the layers. This scenario is consistent with the findings of Refs. \cite{Ji2017, Merkl2019, Zimmermann2021}. Ref. \cite{Zhu2017}, however, reports ultrafast ($< 40\,\mathrm{fs}$) electron transfer from WSe\textsubscript{2} to MoS\textsubscript{2}, independent of the angular alignment of the layers. We speculate that the temporal resolution of the experiment of $190\,\mathrm{fs}$ may have been too poor to resolve the evolution of the electron transfer rate with twist angle.

Finally, in addition to ultrafast charge transfer, ultrafast energy transfer is known to play an important role in vdW heterostructures \cite{Selig2019, Dong2023, Tebbe2024}. In contrast to ultrafast charge transfer that requires the participating layers to hybridize, ultrafast energy transfer is mediated by dipole-dipole interactions. Since the latter are independent of twist angle, ultrafast energy transfer cannot explain the distinct behaviour of WS\textsubscript{2}-graphene heterostructures with twist angles of 0$^{\circ}$ and 30$^{\circ}$ observed in the present manuscript.

\section{Summary and Outlook}

We used trARPES to investigate how the twist angle affects ultrafast charge separation in WS\textsubscript{2}-graphene heterostructures. Charge separation is observed at a twist angle of 0$^{\circ}$ but not at 30$^{\circ}$ where electron and hole transfer proceed at comparable rates. We discussed the role of energy barriers and the tunneling matrix element. A full microscopic understanding, however, requires further theoretical investigations. Our findings provide crucial information for optimizing the performance of vdW heterostructures for specific optoelectronic and photovoltaic applications.

\section{Acknowledgments}
This work received funding from the European Union’s Horizon 2020 research and innovation program under Grant Agreement No. 851280-ERC-2019-STG (DANCE), from the German Science Foundation (DFG) via the Collaborative Research Centre CRC 1277 (Project No. 314695032) and the Research Unit RU 5242 (Project No. 449119662), as well as from the German Federal Ministry of Education and Research (BMBF) (Project No. 05K2022).

\clearpage

\pagebreak

\bibliography{literature_twist_angle}

\pagebreak
	
	\clearpage

	\par\ \ \vspace{0pt}\par\ \ 
	{\centering\bfseries\LARGE Influence of twist angle on ultrafast charge separation in WS\textsubscript{2}-graphene heterostructures\\
		\vspace{20pt}
		SUPPLEMENTAL MATERIAL\\}
	
	\vspace{40pt}
	
	\hrule
	
	\vspace{10mm}

	\section{Fit functions for single- and double-exponential decay}
	
	To extract rise and decay times of a given pump-probe signal, the following analytic fitting function is used that stems from convolving the product of a step function and an exponential decay for the underlying dynamics with a Gaussian function accounting for the finite temporal resolution:
	
	\begin{equation}
		f(t) = \frac{a}{2} \left( 1 + \mathrm{erf}\left( \frac{(t-t_0)\tau - \frac{\text{FWHM}^2}{8\ln2}}{\sqrt{2}\tau\frac{\text{FWHM}}{2\sqrt{2\ln2}}} \right) \right)
		\exp\left( \frac{\frac{\text{FWHM}^2}{8\ln 2}-2(t-t_0)\tau}{2\tau^2} \right)
	\end{equation}
	
	\noindent Here, $a$ is the amplitude, $\text{FWHM}$ is the full width at half maximum of the Gaussian, $t_0$ is the time delay at which pump and probe pulses overlap and $\tau$ is the decay time.\\
	
	\noindent For pump-probe signals that contain two exponentially decaying contributions with different decay times, the following double-exponential decay function is used:
	
	\begin{equation}\begin{split}
			f(t) = \frac{A}{2} \left[a \left(1 + \mathrm{erf}\left(
			\frac{(t - t_0) \tau_1 - \frac{\text{FWHM}^2}{8\ln2}}{\sqrt{2}\tau_1 \frac{\text{FWHM}}{2\sqrt{2\ln2}}}\right)\right) \exp\left( \frac{\frac{\text{FWHM}^2}{8\ln 2}-2(t-t_0)\tau_1}{2\tau_1^2} \right) \right.+\\
			+\left. (1 - a) \left(1 + \mathrm{erf}\left(
			\frac{(t - t_0) \tau_2 - \frac{\text{FWHM}^2}{8\ln2}}{\sqrt{2}\tau_2 \frac{\text{FWHM}}{2\sqrt{2\ln2}}}\right)\right) \exp\left( \frac{\frac{\text{FWHM}^2}{8\ln 2}-2(t-t_0)\tau_2}{2\tau_2^2} \right)\right]
	\end{split}\end{equation}
	
	\noindent Here, $A$ is the main amplitude, $0\leq a\leq1$ is the relative weight of the first decay with respect to the second decay and $\tau_1$ and $\tau_2$ are the decay times.

	\clearpage
	\section{Extracting transient peak positions}
	
	\begin{figure}[h!]
		\centering
		\includegraphics[width=\linewidth]{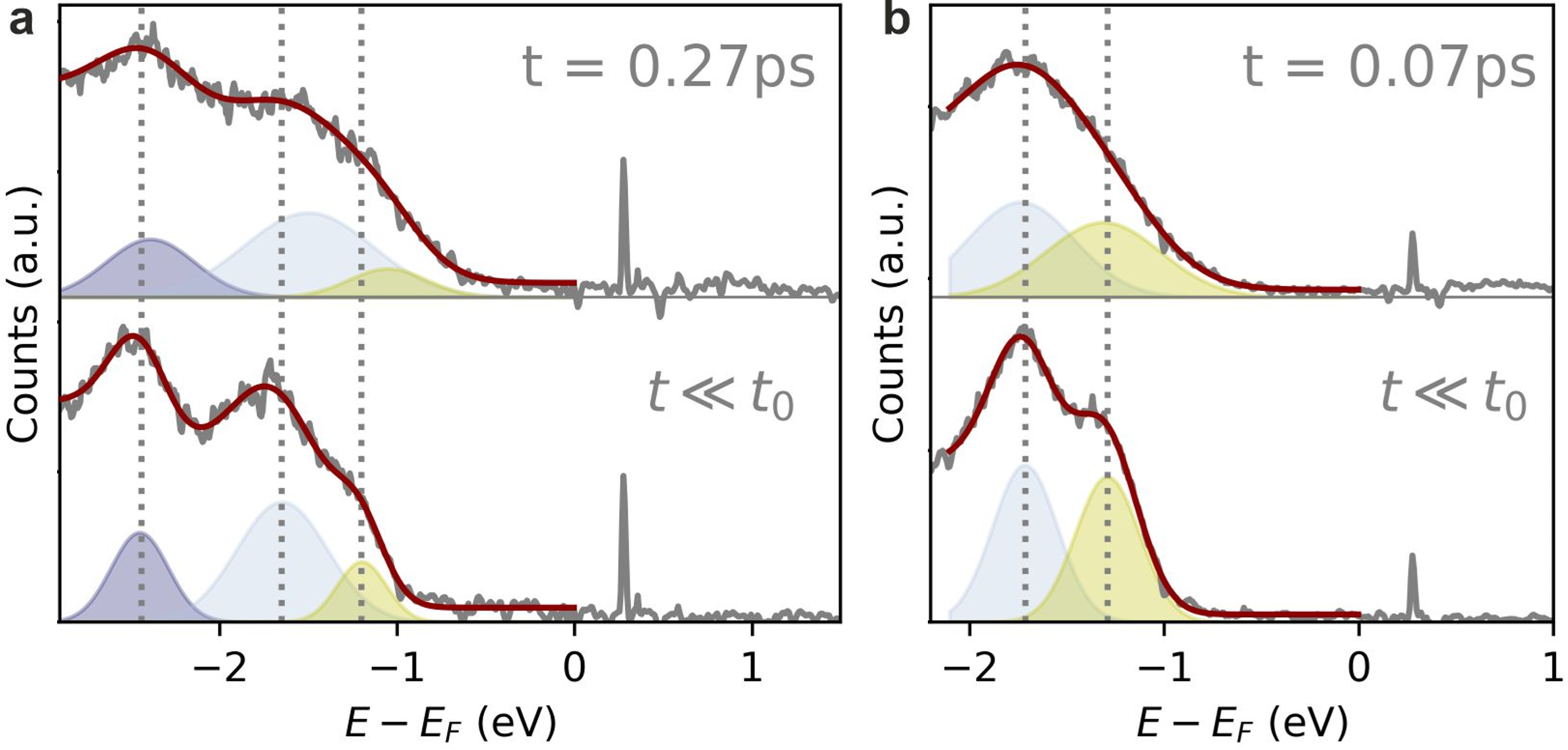}
		\caption{\textbf{Extracting the transient position of the WS\textsubscript{2} valence band. a)} EDCs at the K-point of WS\textsubscript{2} for a twist angle of 0° extracted from Fig.~2a for negative pump-probe delay (bottom) and at $t=\SI{0.27}{\pico\second}$ (top) fitted with a constant background and three Gaussian peaks (red line). The dotted vertical lines mark the peak positions for negative pump-probe delay. The coloured Gaussians illustrate the fit contributions from the three bands. \textbf{b)} Same as a) but for a twist angle of 30°.}
		\label{figureS1}
	\end{figure}
	
	\noindent For extracting the transient position of the WS\textsubscript{2} valence band, the data in Figs. 2a and c (main manuscript) were integrated over the momentum range $\Delta k=\pm\SI{0.06}{\per\angstrom}$ around $k=\SI{1.2}{\per\angstrom}$. The latter was chosen to ensure that the graphene valence band close to the graphene K point does not influence the data. The resulting energy distribution curves (EDCs) were then fitted with a constant background and three Gaussian peaks for the three expected bands. Exemplary fits are presented in Fig.~\ref{figureS1}. The following constraints were applied:
	\begin{itemize}
		\item The constant background was fixed to the value found for negative pump-probe delay.
		\item The energy difference between the two spin-split WS\textsubscript{2} valence bands was fixed to the value found for negative pump-probe delay.
		\item The spectral weight of bands 1 and 2 was fixed to the value found for negative pump-probe delay.
	\end{itemize}
	
\begin{figure}[h!]
		\centering
		\includegraphics[width=0.35\linewidth]{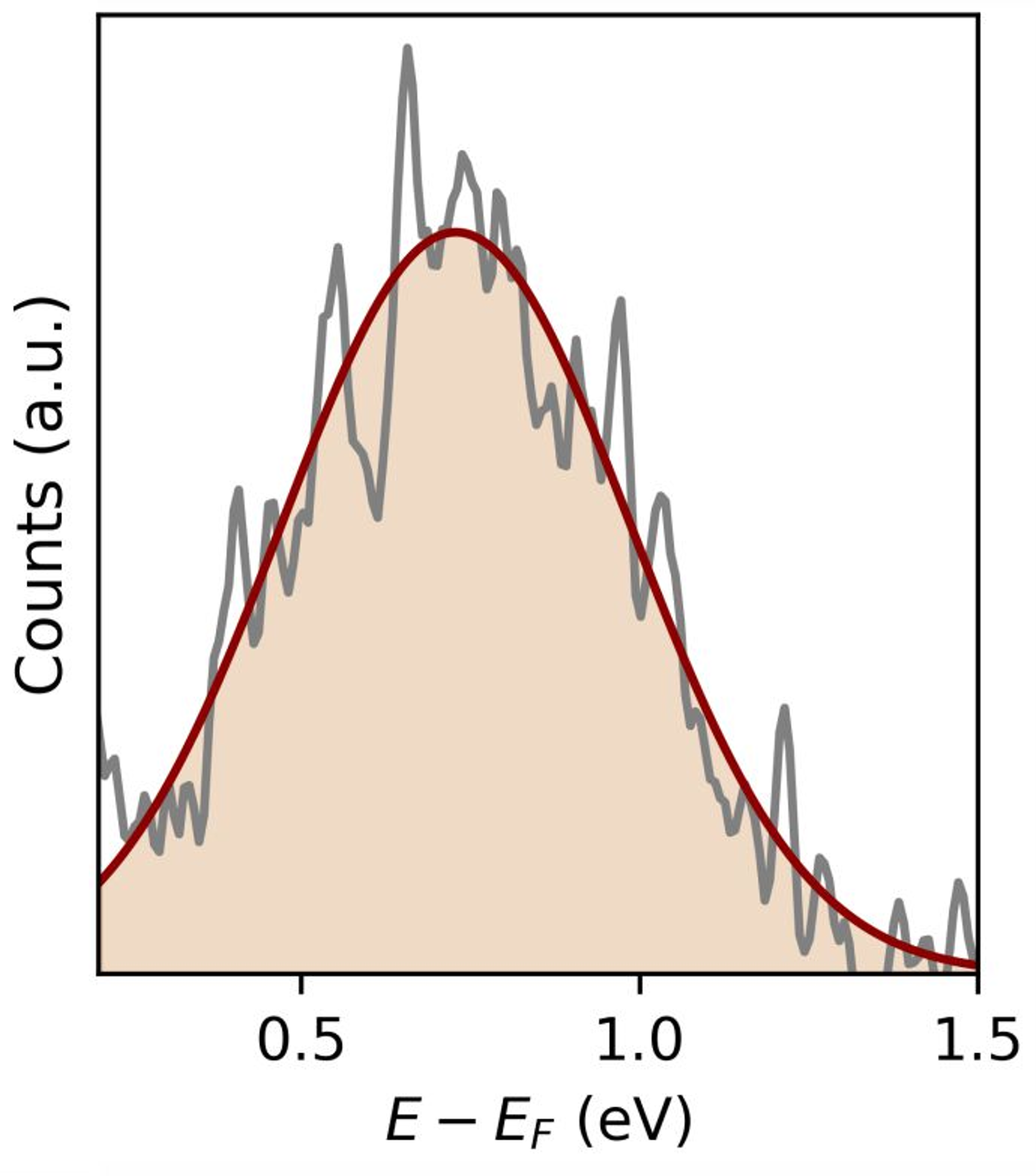}
		\caption{\textbf{Extracting the transient position of the WS\textsubscript{2} conduction band.} EDC at $k=\SI{1.3}{\per\angstrom}$ extracted from Fig.~2d at $t=0.13$\,ps together with Gaussian fit.}
		\label{figureS2}
	\end{figure}

	\noindent To obtain the transient position of the WS\textsubscript{2} conduction band, the data in Figs. 2b and d (main manuscript) were integrated over the momentum range $\Delta k=\pm\SI{0.1}{\per\angstrom}$ around $k=\SI{1.3}{\per\angstrom}$. The resulting EDCs were then fitted with a background-free Gaussian (see Fig.~\ref{figureS2}).	\\
	
	\begin{figure}[h!]
		\centering
		\includegraphics[width=0.65\linewidth]{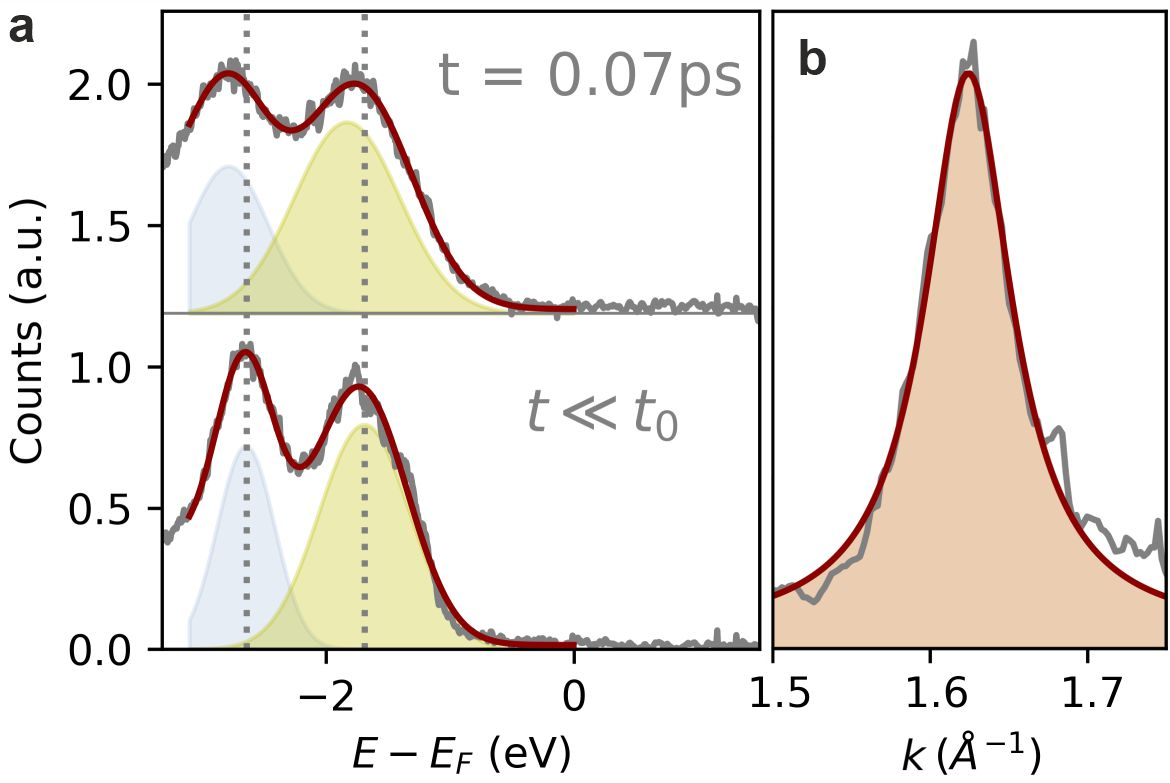}
		\caption{\textbf{Extracting the transient peak positions of the graphene $\pi$-band. a)} EDCs extracted from Fig. 2c at the M point of graphene for negative pump-probe delay and at $t=\SI{0.07}{\pico\second}$ fitted with a Shirley background and two Gaussian peaks (red line). The dotted vertical lines mark the peak positions for negative pump-probe delay. The low-energy peak indicated in light blue corresponds to the graphene valence band. \textbf{b)} MDC extracted from Fig.~2a at $E-E_\mathrm{F}=0.2\,\mathrm{eV}$, fitted with a Lorentzian and a constant offset (red).}
		\label{figureS3}
	\end{figure}
	
	\noindent The shift of the graphene valence band at the graphene M point was determined by integrating the data in Fig. 2c (main manuscript) over the momentum range $\Delta k=\pm\SI{0.05}{\per\angstrom}$ around $k=\SI{1.45}{\per\angstrom}$. The resulting EDCs were fitted with a Shirley background and two Gaussians, where one represents the valence band of WS\textsubscript{2} and the other represents the graphene valence band (see Fig.~\ref{figureS3}a). The spectral weight of the graphene valence band was fixed to the value found for negative pump-probe delay.\\
	
	\noindent The shift of the graphene Dirac cone at the graphene K point was determined by integrating the data in Fig.~2a (main manuscript) over the energy range $\Delta E = \pm\SI{25}{\milli\electronvolt}$ around the five energy positions $(E-E_\mathrm{F})= \SI{-1.0}{\electronvolt},\SI{-0.8}{\electronvolt},\SI{-0.6}{\electronvolt},\SI{-0.4}{\electronvolt},\SI{-0.2}{\electronvolt}$. To improve the signal-to-noise ratio, we grouped the resulting momentum distribution curve (MDC) data into bins of three consecutive time delays. The final MDCs were fitted by the sum of a constant background and a Lorentzian peak (see Fig.~\ref{figureS3}b) for every energy position. The five momentum shifts obtained in this way were averaged and then converted into an energy shift by multiplying with the slope of the $\pi$-band of \SI{7}{\electronvolt\angstrom}.

	\end{document}